\documentclass{PoS}
\usepackage{amssymb}
\usepackage{physics}
\usepackage{graphicx}
\usepackage{comment}
\usepackage{cite}

\usepackage[utf8]{inputenc}
\usepackage{color}

\usepackage{enumitem}
\newlist{myitemize}{enumerate}{10}
\setlist[myitemize]{label*=\arabic*.,nosep,leftmargin=*}

\newcommand\lcdm{$\Lambda$CDM}

\newcommand{\beq}{\begin{equation}}
\newcommand{\eeq}{\end{equation}}
\newcommand{\beqa}{\begin{eqnarray}}
\newcommand{\eeqa}{\end{eqnarray}}

\newcommand{\m}{\mu}
\newcommand{\n}{\nu}

\usepackage{amsmath}

\DeclareMathOperator{\arccosh}{arccosh}

\newcommand{\be}[1]{\begin{equation} \centering \label{#1}} 
\newcommand{\ee}{\end{equation}}
\def\bea{\begin{eqnarray}}
\def\eea{\end{eqnarray}}
\def\ba{\begin{array}}
\def\ea{\end{array}}
\def\bc{\begin{center}}
\def\ec{\end{center}}
\def\bl{\begin{flushleft}}
\def\el{\end{flushleft}}
\def\br{\begin{flushright}}
\def\er{\end{flushright}}
\def\bi{\begin{itemize}}
\def\ei{\end{itemize}}

\def\bt{\begin{tabular}}
\def\et{\end{tabular}}
\newtheorem{question}{Question}
\def\bq{\begin{question}}
\def\eq{\end{question}}
\newtheorem{definition}{Def}
\def\bd{\begin{definition}}
\def\ed{\end{definition}}
\newtheorem{answer}{Answer}
\def\ban{\begin{answer}}
\def\ean{\end{answer}}
\newtheorem{possibleanswer}{Possible answer}
\def\bpa{\begin{possibleanswer}\normalfont}
\def\epa{\end{possibleanswer}}
\newtheorem{theorem}{Theorem}
\def\bth{\begin{theorem}}
\def\eth{\end{theorem}}

\title{Scale symmetry, the Higgs and the cosmos}

\ShortTitle{Scale symmetry, the Higgs and the Cosmos}

\author{\speaker{Javier Rubio}
\\
Department of Physics and Helsinki Institute of Physics, \\  PL 64, FI-00014 University of Helsinki, Finland \\
        E-mail: \email{javier.rubio@helsinki.fi}}

\abstract{I review the Higgs-Dilaton model: a scale-invariant extension of the Standard Model and gravity able to support inflation and dark energy with just an additional degree of freedom on top of the Standard Model content. Potential extensions of the simplest realization on the basis of transverse diffeomorphisms are also discussed.}

\FullConference{Corfu Summer Institute 2019 "School and Workshops on Elementary Particle Physics and Gravity" (CORFU2019)\\
		31 August - 25 September 2019\\
		Corfù, Greece}

\begin{document}
 \vspace*{-5mm}
\begin{flushright}
HIP-2020-8/TH
\end{flushright}
\vspace{.3cm}

\section{Introduction and summary}

The precise measurements of the Higgs mass and the top quark Yukawa coupling at the Large Hadron Collider indicate that we live in a very special universe, at the edge of the absolute stability of the Standard Model (SM) vacuum. This surprising outcome allows us to speculate about consistently extending the SM all the way up to the Planck scale while staying in the perturbative regime. Unfortunately, any sensible attempt to relate the electroweak scale to gravity faces inevitably the so-called \textit{hierarchy problem}, which leads to the instability of the Higgs mass under radiative corrections and to the infamous cosmological constant problem.

The apparent absence of new physics beyond the SM has rejuvenated scale symmetry as an interesting approach to solve the aforementioned hierarchy problem \cite{Shaposhnikov:2008xi,Armillis:2013wya,Gretsch:2013ooa,Chankowski:2014fva,Ghilencea:2015mza,Karananas:2016grc,Ghilencea:2017yqv,Karananas:2017zrg,Shaposhnikov:2018jag,Shaposhnikov:2018nnm,Mooij:2018hew,Ferreira:2018itt,Wetterich:2019qzx,Karananas:2019fox}. A viable scale-invariant theory should exhibit, however, dilatation symmetry breaking in one way or another in order to account for the appearance of physical scales. According to how this symmetry breaking takes place, one can consider two different types of scenarios:
\begin{enumerate}
\item Emergent scale symmetry: Scale symmetry is taken to be anomalous and realized only in the vicinity of non- trivial fixed points \cite{Wetterich:2019qzx,Rubio:2017gty}. The transition among these fixed points happens through a crossover regime where physical scales emerge through dimensional transmutation, as happens for instance in QCD. 
    \item Spontaneously-broken scale symmetry: Scale symmetry is assumed to be exact but spontaneously broken by the non-zero expectation value of a given field or operator. In this type of setting, scale invariance is preserved at the quantum level by means of a scale invariant regularization prescription \cite{Mooij:2018hew}. The price to pay is the lack of renormalizability \cite{Shaposhnikov:2009nk}. 
\end{enumerate}
\noindent In this paper we will focus on the second type of scenarios. Using the so-called \textit{Higgs-Dilaton model} as a proxy, we will describe the cosmological consequences of general biscalar-tensor models displaying a maximally symmetric Einstein–frame kinetic sector and constructed on the basis of scale symmetry and transverse diffeomorphisms (TDiff). The combination of these ingredients leads to some interesting features: 
\begin{enumerate}
    \item Scale-invariant TDiff theories are equivalent to diffeomorphism invariant theories containing a \textit{single} dimensionful parameter $\Lambda_0$.  This quantity appears as an integration constant at the level of the equations of motion, in clear analogy with unimodular gravity scenarios.
    \item For vanishing $\Lambda_0$, one of the two scalar fields becomes the Goldstone boson of dilatations. The combination of gravity and scale invariance forces this massless degree of freedom to have only derivative couplings to matter, evading with it all fifth force constraints.
    \item At large field values, the maximally symmetric structure of the Einstein-frame kinetic sector allows for inflation with the usual slow-roll conditions. The associated observables are almost universal and depend only on the Gaussian curvature of the field-space manifold and the leading-order behavior of the inflationary potential. 
    \item  The presence of a conserved scale current makes the considered models essentially indistinguishable from single–field inflation scenarios, from which they inherit all their virtues. This, in turn, prevents the generation of sizable isocurvature perturbations and non-gaussianities.  
    \item A non-vanishing integration constant $\Lambda_0$ breaks scale symmetry, generating a run-away potential for the dilaton field able to support a dark energy dominated era.  
    \item Beyond its contribution to the late time accelerated expansion of the Universe, the dilaton field remains undetectable by any particle physics experiment or cosmological observation. In particular, and in spite of its ultra-relativistic character, it does not contribute to the effective number of light degrees of freedom  at big bang nucleosynthesis or recombination.
    \item Simple scenarios lead to non-trivial consistency conditions between the inflationary observables and the dark energy equation–of–state parameter, potentially testable with future cosmological observations.
\end{enumerate}

\section{The Higgs-Dilaton model}\label{sec:HDM}

The Higgs-Dilaton model is a natural extension of the SM and gravity where all dimensionfull parameters in the action are replaced by the expectation value of a singlet dilaton field $\chi$.  For the sake of completeness, we will consider two different incarnations of the model: a metric formulation where the connection determining the Ricci scalar is identified with the Levi-Civita connection and a Palatini formulation where the metric and the connection are taken to be independent variables. In both cases, the graviscalar part of the action takes the same form, namely \cite{Shaposhnikov:2008xb,GarciaBellido:2011de}
\beq
\label{eqn:hdm_lagrangian}
S=\int d^4x \sqrt{-g}\left[ \frac{\xi_h h^2 +\xi_\chi \chi^2}{2} g^{\mu\nu} R_{\mu\nu}(\Gamma)
-\frac12 (\partial h)^2- \frac{1}{2} (\partial \chi)^2-V(h,\chi)\right]\,,
\eeq
with $h$ the Higgs field in the unitary gauge,  $\xi_h$ and $\xi_\chi$ two positive-definite constants to be determined from observations, 
\beq\label{eq:f}
V(h,\chi)=\frac{\lambda}{4} \left(h^2-\alpha\chi^2 \right)^2 
\eeq
the SM Higgs potential and $\lambda$ its self-interaction. Note that the direct coupling of the dilaton field to other SM constituents different from the Higgs is forbidden by quantum numbers. 

In this scenario, the vacuum manifold is infinitely degenerate, $h^2=\alpha \chi^2$. The Higgs vacuum expectation value and the Planck mass are generated by the spontaneous symmetry breaking of scale invariance, being the physics independent of the precise dilaton value. Note that this unified mechanism for the generation of masses does not address the  numerical relation among them, making necessary to fine-tune the parameter $\alpha$ in \eqref{eq:f} in order to reproduce the observed hierarchy of scales ($\alpha\sim v^2/M_P^2\sim 10^{-32}$), see, however, Refs.~\cite{Karananas:2016grc,Shaposhnikov:2018jag,Shaposhnikov:2018xkv}. 

The cosmological consequences of the Higgs-Dilaton model \eqref{eqn:hdm_lagrangian} are more easily understood in the so-called Einstein frame in which the non-linearities associated with the non-minimal couplings to gravity are moved to the scalar sector of the theory. This frame is obtained by performing a Weyl transformation $g_{\mu\nu}\to \Omega^2 g_{\mu\nu}$ with conformal factor $\Omega^{2}\equiv (\xi_h h^2+\xi_\chi \chi^2)/M_P^2$.  While  this field redefinition modifies the Ricci scalar in the metric formulation, it leaves it invariant in the Palatini one since the metric and the connection are in this case unrelated. This translates into a different Einstein-frame action for the metric and Palatini cases, namely
\beq\label{SE1}
S=\int d^4x \sqrt{-g}\left[\frac{M_P^2}{2} R - 
\frac{1}{2} g^{\mu\nu} \gamma_{ab} \partial_\mu \varphi^a \partial_\nu \varphi^b - U_(\varphi^a)\right]\,,
\eeq
with  $R = g^{\mu\nu}R_{\mu\nu}(\Gamma)$ the Ricci scalar,\footnote{Note that in this frame the gravitational part of the action takes the usual Einstein-Hilbert form, allowing to identify the connection with the standard Levi-Civita connection both in the metric and Palatini formulations.} $\varphi^{a}=(\varphi^1,\varphi^2)=(h,\chi)$ a convenient notation identifying the fields as coordinates in a 2-dimensional field-space $\cal M$, $U(\varphi^a)\equiv V(h)/ \Omega^{4}(h,\chi)$ a Weyl-rescaled potential and
\beq
  \gamma_{ab} = \frac{1}{\Omega^2}\left( \delta_{ab} + y \times \frac{3}{2} M_P^2 
  \frac{\partial_a \Omega^2 \partial_b \Omega^2}{\Omega^2}\right)
\eeq 
the metric in $\cal M$ with $y=1$ for the metric formulation and  $y=0$ for the Palatini one. 

A simple computation of the Gauss curvature associated with the field-space metric $\gamma_{ab}$ reveals that it is not generally possible to canonically normalize both fields by performing additional field redefinitions \cite{GarciaBellido:2011de}. An important simplification can be obtained, however, by considering the Noether current of scale symmetry. For homogeneous fields in a Friedmann-Lemaitre-Robertson-Walker background, the conservation equation for this current takes the simple form \cite{GarciaBellido:2011de}
\begin{equation}\label{eq:conserv}
\frac{1}{a^3}\frac{d}{d t}\left(a^3 \gamma_{ab}\dot\varphi^a\Delta\varphi^b\right)=0\,,
\end{equation}
with $a = a(t)$ the scale factor, the dots denoting derivatives with respect to the coordinate time $t$ and $\Delta\varphi^a$ the infinitesimal action of dilatations on the fields. Taking into account the explicit form of $\Delta\varphi^a$ in the Einstein-frame, we can rewrite Eq.~\eqref{eq:conserv} as
\begin{equation}\label{eq:conserv2}
\dot \phi+3H\phi=0\,,
\end{equation}  
with
\begin{equation}\label{phidef}
\phi^2(h,\chi)\equiv  (1+6\,y\,\xi_h)h^2+ (1+6\,y\,\xi_\chi)\chi^2 
\end{equation}
a \textit{radial field variable} in the $\lbrace h,\chi\rbrace$ plane \cite{GarciaBellido:2011de}.
\begin{figure}[!t]
\begin{center}
	\includegraphics[width=0.55\textwidth, trim={0 5cm 0 5cm},clip]{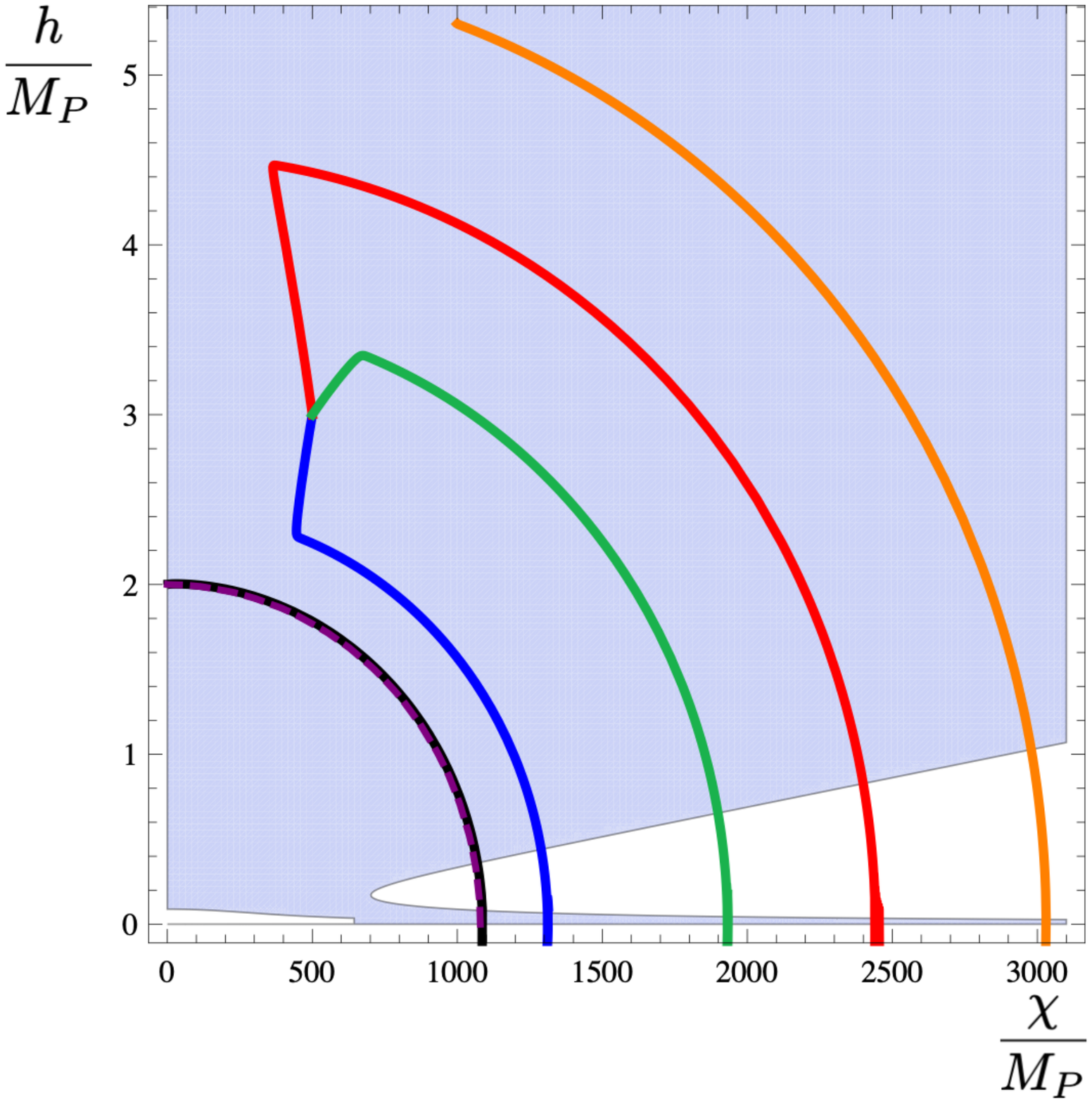}
	\caption{\label{fig:traj} Numerically computed trajectories for the Higgs and dilaton fields in the Einstein--frame.
	Although these trajectories are initially affected by the precise initial conditions (slow-roll for the lower and upper curves and non slow-roll for the intermediate ones), the fields end always in an elliptical trajectory of constant radius $\phi$.}
	\end{center}
\end{figure}
Independently of any assumption on the initial conditions (cf.~Fig.~\ref{fig:traj}), Eq.~\eqref{eq:conserv2} guarantees the evolution of \eqref{phidef} to a constant value in an expanding Universe \cite{GarciaBellido:2011de} (see also Refs.~\cite{Ferreira:2016vsc,Ferreira:2016wem,Ferreira:2018itt,Ferreira:2018qss}).  The existence of this attractor solution has important phenomenological consequences.  To see this explicitly, let us perform a field redefinition involving the \textit{radial field} $\phi$ and a complementary \textit{angular variable} $\Theta$ invariant under the simultaneous rescaling of $h$ and $\chi$, namely 
\begin{equation}
\exp\left[\frac{\gamma\, \Phi}{M_P}\right] \equiv \sqrt{\frac{\kappa_c}{\kappa}} \frac{\phi}{M_P}\,, \hspace{20mm}
\gamma^{-2} \Theta \equiv \frac{\phi^2}{\Omega^2 M_P^2}\,,\label{eq:Thetadef} 
\end{equation}
with
\beq\label{kappadef}
\kappa_c \equiv-\frac{\xi_h}{1+6\,y\, \xi_h}\,, \hspace{15mm} 
\kappa \equiv\kappa_c\left(1-\frac{\xi_\chi}{\xi_h}\right)\,,\hspace{15mm}
\gamma \equiv \sqrt{\frac{\xi_\chi}{1+6 \,y\, \xi_\chi}}\,.
\eeq 
In terms of these variables, the Einstein-frame action \eqref{SE1} takes the very compact form \cite{Karananas:2016kyt,Casas:2017wjh}
\begin{equation}\label{action_HD2}
S=\int d^4x \sqrt{-g}\left[ \frac{M_P^2}{2}R 
-\frac{K(\Theta)}{2}(\partial\Theta)^{2} 
- \frac{\Theta}{2}(\partial \Phi)^2 
 -U(\Theta)\right]\,,
\end{equation}
with 
\beq\label{Utheta}
K(\Theta)=-\frac{M_P^2}{4\, \Theta}\left(\frac{1}{\kappa\Theta+c}+\frac{a}{1-\Theta}\right)\,, \hspace{15mm} U(\Theta)= U_0(1-\Theta)^2\,, 
\eeq
and 
\beqa\label{cdef}
U_0\equiv\frac{\lambda \, a^2 M_P^4}{4}\,, \,\hspace{15mm} 
a \equiv \frac{1+6 \,y\, \kappa}{\kappa}\,, \, \hspace{15mm} 
c\equiv\frac{\kappa}{\kappa_c}\gamma^2\,.
\eeqa
As clearly seen in \eqref{action_HD2}, the \textit{rescaled radial field} $\Phi$ appears only through derivatives, not generating therefore any fifth-force effect among the SM constituents \cite{GarciaBellido:2011de} (see also Refs.~\cite{Ferreira:2016kxi,Burrage:2018dvt}). This emergent shift symmetry can be understood as a non-linear realization of the original dilatation symmetry in Eq.~\eqref{eqn:hdm_lagrangian}, with $\Phi$ the associated Goldstone boson or dilaton.

\subsection{Inflation}

The freezing of $\Phi$ as the Universe expands confines the motion of the fields to ellipsoidal trajectories in the $\lbrace h,\chi\rbrace$ plane. This means that, in spite of dealing with an intrinsically two-field model, the dynamics of the system turns out to be essentially single field, preventing the generation of large isocurvature perturbations or non-Gaussianities \cite{GarciaBellido:2011de} (see also Ref.~\cite{Ferreira:2018qss}). For all practical purposes, inflation is driven by the \textit{angular field} $\Theta$, which intentionally displays a non-canonical normalization with a remarkable pole structure. The pole at $\Theta=1$ in Eq.~\eqref{action_HD2} is a \textit{Minkowski pole}, around which the usual SM minimally coupled to gravity is approximately recovered. For the purposes of inflation, this pole can be safely neglected, leaving behind an approximate  action 
\begin{equation}\label{action_HD3}
S=\int d^4x \sqrt{-g}\left[ \frac{M_P^2}{2}R-
\frac{1}{2}\Bigg(-\frac{M_P^2(\partial \Theta)^{2}}{4\, \Theta (\kappa
\Theta+c)}+ \Theta(\partial \Phi)^2 \Bigg) -U(\Theta)\right]\,,
\end{equation}
with a \textit{maximally symmetric kinetic sector of negative curvature $\kappa$} \cite{Karananas:2016kyt,Casas:2017wjh}. The remaining poles at $\Theta=0$ and $\Theta=-c/\kappa$ lead to an effective stretching of the canonically normalized variable
\begin{equation}\label{eq:cantheta0}
\theta=\int^\theta \frac{d\Theta}{\sqrt{-4\,  \Theta
(\kappa \Theta+c)}}\,,
\end{equation}
and the associated flattening of the potential around them \cite{Artymowski:2016pjz}.  For $\xi_\chi=0$ ($c=0$), the pole $\Theta=0$ becomes quadratic and one recovers the well-known exponential stretching of Higgs inflation models \cite{Rubio:2018ogq,Tenkanen:2020dge}
\begin{equation}\label{eq:canonicalpotHI}
U(\theta)=U_0\left(1-e^{-2\sqrt{-\kappa}\vert \theta\vert/M_{\rm P}}\right)^2\,.
\end{equation}
For non-vanishing values of $\xi_\chi$ ($c\neq 0$), the inflationary pole at $\Theta=0$ is no longer reachable and we are left with a linear pole at $\Theta=-c/\kappa$. In this case, the stretching of the potential is restricted to a compact field range,
\begin{equation}\label{eq:canonicalpot}
U(\theta)=U_0\left(1+\frac{c}{\kappa}\cosh^2 \frac{\sqrt{-\kappa}(\theta_0-\vert \theta\vert)}{M_{\rm P}}\right)^2\,, 
\end{equation}
with $\theta_0\equiv -M_{\rm P}/\kappa\, \arccosh (\sqrt{-\kappa/c})$ an integration constant shifting to zero the location of the potential minimum in order to facilitate the comparison with Eq.~\eqref{eq:canonicalpotHI}.  

Since we are especially interested in a scenario in which the dilaton field plays a subdominant role during inflation ($\xi_{\chi}\ll \xi_{h}$), we can approximate $\vert \kappa\vert\simeq \vert\kappa_c\vert$ and assume the ratio $c/\vert \kappa\vert$ to be small.  Following the standard procedure for computing the inflationary observables in the slow-roll approximation, we can easily obtain the following analytical expressions for the amplitude, the tilt and the running of primordial power spectrum of curvature fluctuations~\cite{Casas:2017wjh,Almeida:2018oid}, 
\begin{equation}\label{As}  
A_s
 = \frac{1}{6 a^2 \vert\kappa_c\vert} \frac{\lambda_{s} \sinh^{2} \left(4 c N_* \right)}{1152\pi^2 \,  c^2} \,,\hspace{10mm}  n_s = 1-8\,c \coth\left(4 c N_*\right)\,,   \hspace{10mm}  \alpha_s = - 32\, c^2 \sinh^{-2} \left(4 c N_* \right)\,,
\end{equation}
as well as the tensor-to-scalar ratio
\begin{equation}\label{nsr2}
r=  \frac{32\, c^2}{\vert \kappa_c\vert} \sinh^{-2} \left(4 c N_*\right)\,.
\end{equation}
The number of e-folds $N_*$ to be inserted in these expressions depends on the details of the heating stage, which, up to order ${\cal O}(\xi_\chi/\xi_h)$ corrections and a negligible production of dilaton particles \cite{GarciaBellido:2012zu}, coincides with that of Higgs inflation \cite{Bezrukov:2008ut,GarciaBellido:2008ab,Repond:2016sol,Rubio:2019ypq}.  As shown in these references, the depletion of the inflaton condensate takes place rather fast, strongly constraining the number of $e$-folds of inflation to the narrow interval $60 \lesssim N_*\lesssim 62$ \cite{Casas:2017wjh}. Given this window and the value of the Higgs self-coupling following from the evolution of the renormalization group equations up to the inflationary scale\footnote{See Refs.~\cite{Bezrukov:2014ipa,Rubio:2014wta,Bezrukov:2017dyv} for a discussion of the subtleties associated with this procedure.} we can determine the non-minimal couplings $\xi_h$ and $\xi_\chi$ by comparing the above inflationary predictions to CMB data.  As we will see below, this apparently trivial parameter-fixing has a strong impact in the subsequent evolution of the Universe.

\subsection{Dark energy}\label{sec:DE}

Once reheating is complete, the system relaxes to the minimum of the potential, with the final expectation value of $\Theta$  determined by the small parameter $\alpha$ we have neglected so far. From there on, the evolution of the Universe proceeds in the usual way, undergoing standard radiation- and matter-dominated eras \cite{GarciaBellido:2011de}. A natural question at this point is how to recover the present accelerated expansion of the Universe without  spoiling the scale symmetry of the system in an uncontrollable  way. A simple possibility advocated in Refs.~\cite{Shaposhnikov:2008xb,GarciaBellido:2011de} is to replace General Relativity by Unimodular Gravity \cite{vanderBij:1981ym,Buchmuller:1988wx,Unruh:1988in}. Note that this replacement does not constitute a strong change of paradigm since both theories are special cases within the minimal group of transverse diffeomorphisms required for having spin-two gravitons \cite{vanderBij:1981ym,Buchmuller:1988wx,Alvarez:2006uu} (cf. Section \ref{sec:beyond}). 

In unimodular gravity, the metric determinant is restricted to take a constant value $g = -1$, forbidding the inclusion of a cosmological constant in the action. The ``cosmological constant" reappears, 
however, as a $\Lambda_0$ integration constant at the level of the equations of motion \cite{Shaposhnikov:2008xb,GarciaBellido:2011de}. In the presence of non-minimal couplings to gravity, this constant becomes the strength of a potential for the dilaton field $\Phi$ when moving to the Einstein frame, namely  \cite{Shaposhnikov:2008xb,GarciaBellido:2011de}
\begin{eqnarray}\label{potentialL}
V_\Lambda=\frac{{\Lambda_0}}{c^2} e^{-\frac{4\gamma\Phi}{M_P}}\,,
\end{eqnarray}
with $\gamma$ and $c$ given in Eqs.~\eqref{kappadef} and \eqref{cdef}, respectively.
This exponential potential breaks down the scale symmetry of the original theory in a controllable way, allowing to support an accelerated expansion of the Universe for positive $\Lambda_0$ values while driving the scalar field to infinity \cite{Wetterich:1987fk,Wetterich:1987fm}. The radial variable $\Phi$ behaves essentially as a thawing quintessence field \cite{Ferreira:1997hj,Caldwell:2005tm} and stays frozen until the decreasing energy density of the matter sector equals the approximately constant energy density in the potential term \eqref{potentialL}. When that happens, the Universe enters an accelerated expansion era characterized by an equation-of-state parameter \cite{Scherrer:2007pu} 
\beq \label{eqn:eos_omega_quintessence}
w=\frac{16 \gamma^2}{3} F(\Omega_{\rm DE})-1\,, 
\eeq
with  $\Omega_{\rm DE}$ the dark-energy/dilaton fraction and  $F(\Omega_{\rm DE})$ a monotonically increasing function smoothly interpolating between $F(0)=0$ in the deep radiation and matter dominated eras and $F(1)=1$ in the asymptotic dark--energy dominated era (for details, see Ref.~\cite{Casas:2017wjh}).  Combining this expression with Eqs.~\eqref{As} and \eqref{nsr2} brings us to an unusual situation in which the early and late Universe observables, customarily understood as independent parameters, become related in a rather non-trivial manner \cite{GarciaBellido:2011de,Casas:2017wjh}, namely
\begin{equation}\label{nswcons}
n_s = 1- \frac{2}{N_*}X \coth X\,,\hspace{12mm} r=\frac{2}{\vert\kappa_c \vert N_*^2}  X^2\sinh^{-2} X\,, \hspace{12mm} X\equiv 4 c N_*=\frac{3 N_*(1+w)}{4 F(\Omega_\textrm{DE})}\,.
\end{equation}
Note that, although we have restricted restricted ourselves to a classical treatment, these  consistency relations remain valid even in the presence of quantum corrections, provided that the ultraviolet completion of the SM non-minimally coupled to gravity respects the scale symmetry of the tree--level action \cite{Bezrukov:2012hx}. Significant deviations from them are only expected for very fine-tuned regions in parameter space leading to the appearance of an inflection point along the inflationary trajectory \cite{Rubio:2014wta,Bezrukov:2017dyv}.

\section{Data comparison}

\begin{figure}[!t]
 \includegraphics[width=\textwidth]{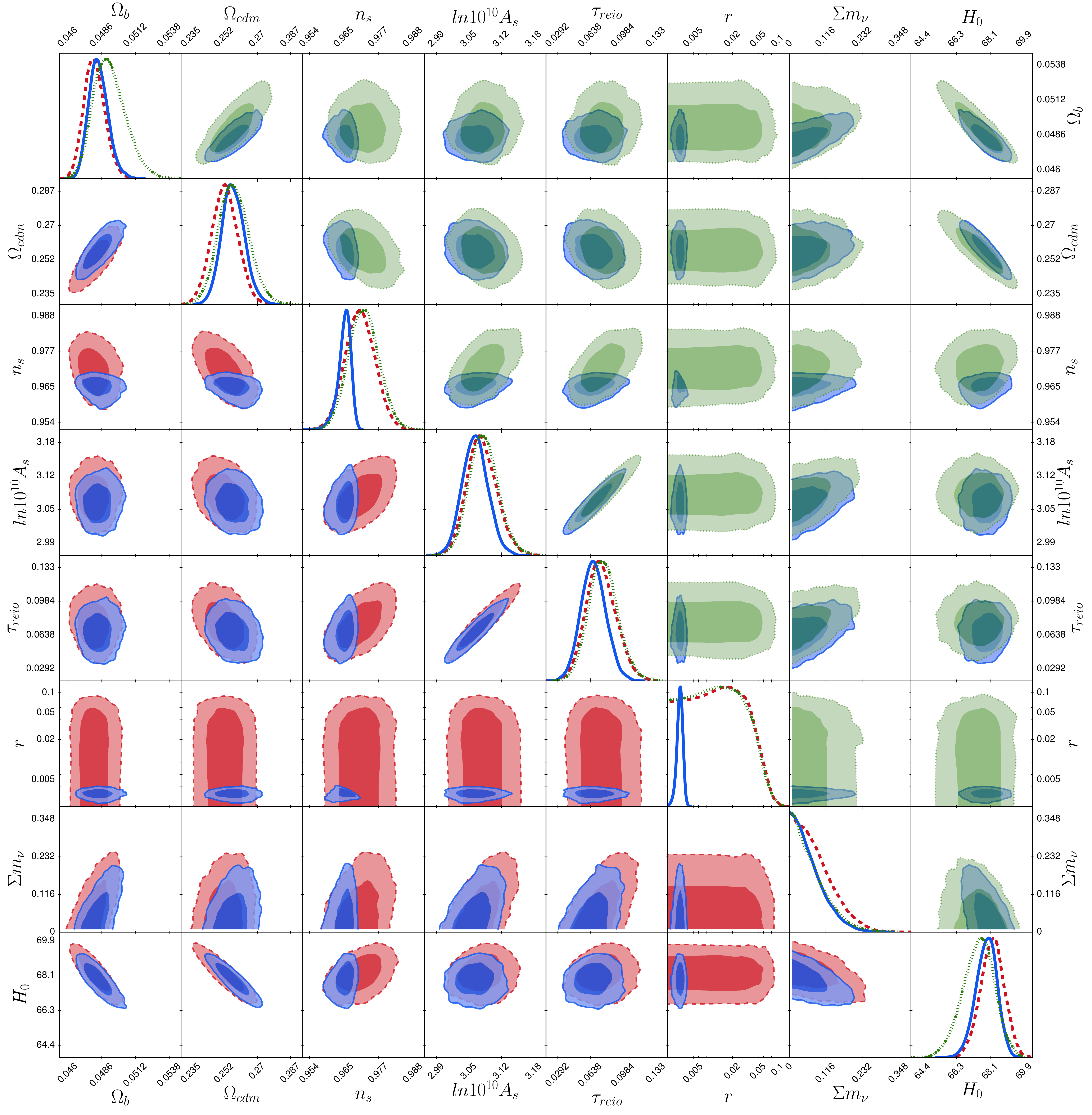}
 \caption{\label{fig:mcmc} Comparison of the MCMC results for the $\Lambda$CDM model (red, dashed), the HD model (blue, solid) and $w$CDM (green, dotted). Data sets include the 2015 Planck high--multipole $TT$ likelihood, the 2015 Planck low--multipole polarization and temperature likelihoods, the 2015 Keck/Bicep2 likelihood data release, the Joint Lightcurve Analysis data and the baryon acoustic oscillation data from 6dF, BOSS LOWZ, BOSS CMASS and SDSS. Taken from Ref.~\cite{Casas:2017wjh}.}
\end{figure}

 To understand the impact of the conditions \eqref{nswcons} on cosmological observables we compare the Higgs-Dilaton model with a standard $\Lambda$CDM scenario and a $w$CDM model involving also a thawing quintessence potential but without any additional constraints on the primordial power spectra \cite{Casas:2017wjh} (see also Ref.~\cite{Trashorras:2016azl}). The posterior probability distributions following from a MCMC analysis of these three models are shown in Fig.~\ref{fig:mcmc}. 
As clearly appreciated in this plot, the performance of the Higgs-Dilaton model as compared to the other two scenarios is certainly remarkable. Indeed, as confirmed by the explicit determination of the Bayes factor $B$ \cite{Casas:2017wjh}, the Higgs-Dilaton model is \textit{strongly preferred} with respect to  a quintessence scenario without consistency relations ($\ln\, B=3.51$),  while displaying a \textit{positive evidence} over $\Lambda$CDM ($\ln \,B=0.88$) \cite{Casas:2017wjh}.~\footnote{According to the so-called Kass and Raftery scale \cite{Kass:1995loi}, a value $\vert \Delta \ln \,B \vert>3$ is understood as a strong statistical preference.} 

\begin{figure}[!t]
 \includegraphics[width=\textwidth,,trim={0cm 0 0 7cm },clip]{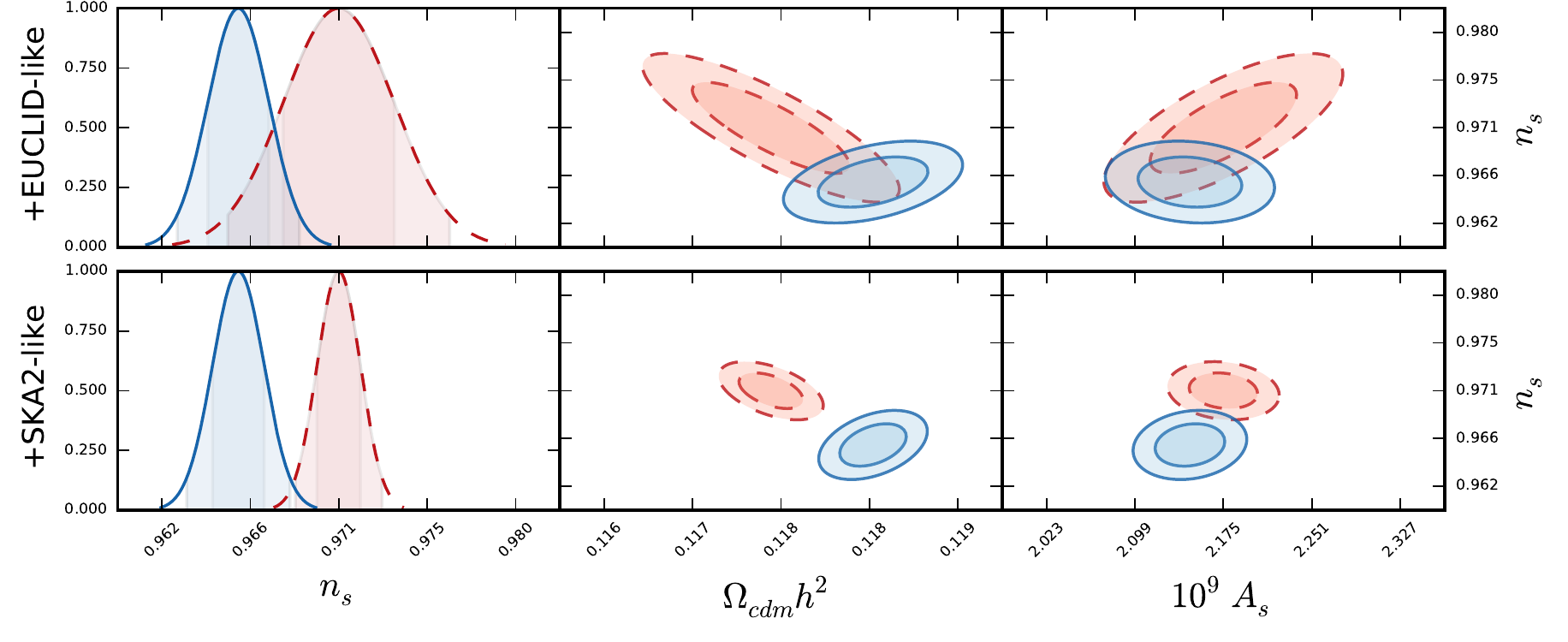}
 \caption{\label{fig:corrMats1} Forecast comparison of the 1D Gaussian probability distribution for the spectral tilt (first column) and the $1\sigma$ and $2\sigma$ 2D Gaussian error contours (second and third columns) for Higgs-Dilaton (blue, solid) and \lcdm{} cosmologies (red, dashed), centered on the fiducial values obtained from \textit{its own} MCMC run. The two rows illustrate the expected parameter restrictions following from i) a Euclid-like mission combining galaxy clustering on linear scales and non-linear weak lensing and ii) a SKA2-like survey with both non-linear galaxy clustering and weak lensing. Adapted from Ref.~\cite{Casas:2017wjh}}
\end{figure}

Given the mild dominance of the Higgs-Dilaton model over $\Lambda$CDM, it is tempting to  estimate  whether future galaxy clustering and weak lensing probes will
be able to discriminate between these two well-motivated scenarios. As shown in Fig.~\ref{fig:corrMats1}, the combination of present data sets with the forecasted sensitivities of Euclid or SKA2 missions could translate into a more than 3$\sigma$ separation between the Higgs-Dilaton model and $\Lambda$CDM \cite{Casas:2017wjh} (provided, of course, that future best fits values remain comparable to the present ones). This interesting result could be significantly improved by future polarization experiments such as the Simon Observatory and the LiteBIRD satellite, expected to determine the tensor-to-scalar ratio at the level of 1 part in a 1000.

\begin{figure}[!t]
\begin{center}
	\includegraphics[width=0.7\textwidth]{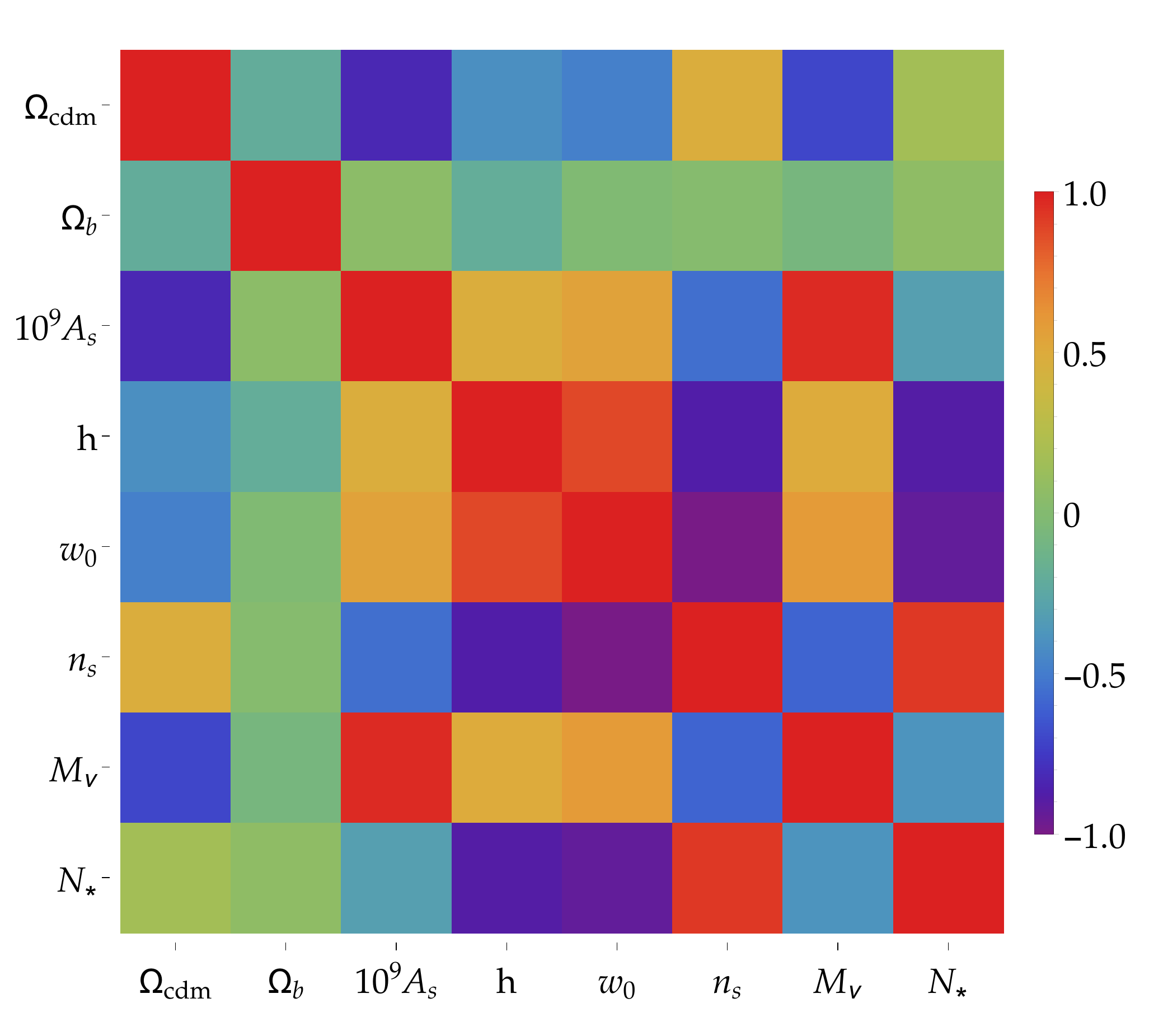}
	\caption{\label{fig:corrmat_HD_Euc} Higgs-Dilaton correlation matrix for a Euclid-like probe, with the $+1$ and $-1$ limits standing respectively for \textit{total correlation} and \textit{anti-correlation}. Taken from Ref.~\cite{Casas:2017wjh}.}
	\end{center}
\end{figure}

It i also interesting to notice that the impact of the consistency relations \eqref{nswcons} extend to other cosmological parameters not directly related to inflation or dark energy. This is partially reflected in Fig.~\ref{fig:corrMats1}, and more accurately illustrated in Fig.~\ref{fig:corrmat_HD_Euc}, where we display the Higgs-Dilaton correlation matrix for a Euclid-like probe. Among other minor changes, we observe that characteristic \lcdm{} features, such as the negative correlation between the spectral tilt $n_s$ and the dark matter fraction $\Omega_{cdm}$ or the positive correlation between the reduced Hubble rate $h$ and the present equation-of-state parameter $w_0$, become significantly altered in the Higgs-Dilaton scenario. 

\section{Beyond the Higgs-Dilaton model}\label{sec:beyond}

As we have seen in the previous section, the main ingredients of the Higgs-Dilaton model are i) scale symmetry, ii) the existence of a singlet scalar field on top of the SM content and iii) unimodular gravity. Is it possible to unify all these pieces into a common framework? To answer this question, it is convenient to go back to the minimal group of transverse or volume-preserving diffeomorphisms mentioned in Section \ref{sec:DE}. This subgroup of coordinate transformations is generated by the subalgebra of transverse vectors 
\begin{equation}\label{tdiff_def}
x^\mu\to x^\mu+\xi^\mu\,, \hspace{20mm}  \partial_\mu \xi^\mu=0\,,
\end{equation}
and, contrary to what happens in diffeomorphism--invariant theories, does not completely determine the form of the action. In particular, it is always possible to include arbitrary functions of the metric
determinant in front of all terms, since this quantity transforms as a scalar under volume--preserving diffeomorphisms and generically becomes a propagating degree of freedom. Having this in mind, let us can consider the most general TDiff action invariant under dilatations and containing at most two field derivatives. This is given by \cite{Blas:2011ac}
\begin{equation}
\begin{aligned}
\label{Tdiffinit}
S=\int d^4x \sqrt{-g}\Bigg\{ \frac{\phi^2f(g)}{2} R&-
\frac{\phi^2}{2}\Big[G_{1}(g) \left(\partial g\right)^2-2 \,G_{2}(g)
\left(\partial g\right)\left(\frac{\partial\phi}{\phi} \right)
+G_{3}(g) \left(\frac{\partial \phi}{\phi} \right)^2 \Big] -
\phi^4v(g)\Bigg\}\,,
\end{aligned}
\end{equation}
with $f$, $G_1$, $G_2$, $G_3$ and $v$ arbitrary \textit{theory-defining functions} of the metric determinant $g$.  The existence of an additional scalar degree of freedom in \eqref{Tdiffinit} can be made explicit by reformulating it in a diffeomorphism invariant way.  
This can be done by performing a \textit{general} coordinate transformation with Jacobian $J(x)\neq 1$ followed by the introduction of a St\"uckelberg compensator field $a(x)\equiv J(x)^{-2}$ transforming under diffeomorphisms as the metric determinant, i.e. $\delta_\xi a =\xi^\mu \partial_\mu a +2 a \partial_\mu \xi^\mu$. We obtain~\cite{Blas:2011ac,Karananas:2016grc} 
\begin{equation}
\begin{aligned}
\label{eqn:hdm_lagrangianL} 
S=\int d^4x \sqrt{-g}\Bigg\{ \frac{\phi^2f(\Theta) }{2}
R&-\frac{\phi^2}{2}\Big[G_{1}(\Theta) (\partial
\Theta)^2+2 \,G_{2}(\Theta) (\partial
\Theta)\left(\frac{\partial\phi}{\phi} \right)+G_{3}(\Theta)
\left(\frac{\partial\phi}{\phi}\right)^2 \Big] \\ &-\phi^4v(\Theta)
-\frac{\Lambda_0}{\sqrt{\Theta}} \Bigg\}\,,
\end{aligned}
\end{equation}
with $\Theta\equiv g/a>0$ and $\Lambda_0$ a \textit{unique} symmetry--breaking scale that arises as an integration constant in the original TDiff formulation.~\footnote{For details on this point, the reader is referred to Ref.~\cite{Blas:2011ac}.} Written this way, the theory is manifestly invariant under the internal transformations $g_{\m\n}(x)\mapsto \lambda^2 g_{\m\n}(x)$, $\phi(x) \mapsto \lambda\,\phi(x)$, $ \Theta(x)\mapsto \Theta(x)$ if $\Lambda_0=0$. 

As we did in the Higgs-Dilaton scenario it is convenient to recast the action \eqref{eqn:hdm_lagrangianL} in an Einstein-frame form. Assuming the prefactor of the Ricci scalar to be positive-definite ($\phi^2f(\Theta) > 0$) and performing a Weyl rescaling of the metric 
$g_{\mu\nu}\to M_P^2/(\phi^2f(\Theta))\, g_{\mu\nu}$ together  by a field redefinition~\cite{Blas:2011ac,Karananas:2016kyt}
\begin{equation}
\Phi= M_P\ln\left(\frac{\phi}{M_P}\right)  - M_P \int^{\Theta}
d\Theta\, \frac{K_{2}(\Theta)}
{K_{3}(\Theta)}\,,
\end{equation}
we obtain
\begin{equation}\label{eq:sigmamodel}
S=\int d^4x \sqrt{-g}\Bigg\{ \frac{M_P^2}{2}R -\frac{1}{2}\Bigg[ 
K(\Theta)(\partial \Theta)^2+K_3(\Theta) (\partial
\Phi)^2 \Bigg]-U(\Theta)-U_{\Lambda_0}(\Theta,\Phi)\Bigg\}\,,
\end{equation}
with 
\begin{equation}
K(\Theta)=M_P^2 \left(\frac{K_1(\Theta)
K_3(\Theta)-K_2^2(\Theta)}{K_3(\Theta)}\right)\,,
\end{equation}
\begin{equation}
K_1(\Theta)\equiv\dfrac{G_1(\Theta)}{f(\Theta)}+
\dfrac{3}{2}\left(\dfrac{f'(\Theta)}{f(\Theta)}\right)^2\
,\hspace{8mm}
K_2(\Theta)\equiv\dfrac{G_2(\Theta)}{f(\Theta)}+3\,
\dfrac{f'(\Theta)}{f(\Theta)}\ , \hspace{8mm}
K_3(\Theta)\equiv
6+\dfrac{G_3(\Theta)}{f(\Theta)}\,,
\end{equation}
the primes denoting derivatives with respect to $\Theta$ and 
\begin{equation}
U(\Theta)=\frac{M_P^4\,  v(\Theta)}{f^2(\Theta)}\,,
\hspace{8mm}  U_{\Lambda_0}(\Theta,\Phi)=\Lambda_0\, K_\Lambda(\tilde \theta)
\,e^{-4 \Phi/M_P } \,, \hspace{8mm}
K_\Lambda=
\frac{\exp \left({4
\int\,\frac{K_2(\Theta)}{K_3(\Theta) }}\, d
\Theta \right)}{f^2(\Theta)\sqrt{\Theta}} \ .
\end{equation}
\begin{figure}[!t]
\begin{center}
	\includegraphics[width=0.6\textwidth, trim={0 0.3cm 0 0},clip]{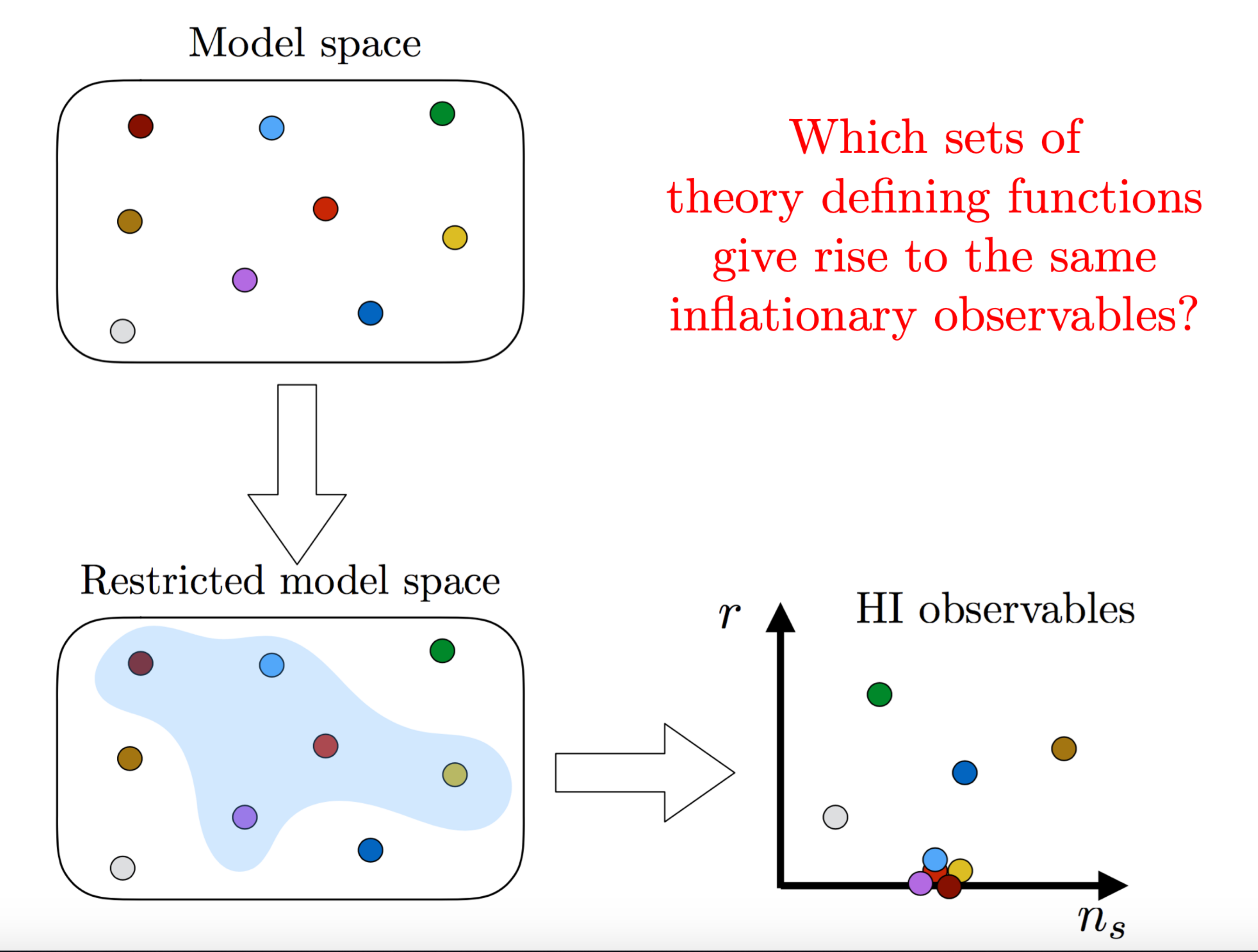}
	\caption{\label{fig:modelsel} Maximally symmetric \textit{universality class}. }
	\end{center}
\end{figure}
Given this general action, one can construct a \textit{preferred class of scale-invariant models} by imposing sensible physical conditions such as the absence of ghosts ($K(\Theta) > 0$,\, $K_3(\Theta) > 0$) or anti-de Sitter regimes ($U(\Theta)\geq 0$,\, $\Lambda_0\, K_\Lambda(\Theta)\geq 0$). On top of that, one could account for phenomenological aspects such as the recovery of SM action at low-energies, the non-generation of massless excitations excessively contributing to the effective number of relativistic degrees of freedom at Big Bang Nucleosynthesis and recombination, or  the existence of inflationary solutions compatible with observations. An \textit{equivalence class of theories} satisfying the last two requirements can be easily constructed by requiring the kinetic manifold in Eq.~\eqref{eq:sigmamodel} to be maximally symmetric (cf.~Fig~\ref{fig:modelsel}) \cite{Karananas:2016kyt, Casas:2018fum}. In this case, the differential equation satisfied by the Gaussian curvature 
\begin{equation}
\kappa(\Theta)=\frac{K_3'(\Theta)F'(\Theta)-2F(\Theta) K''_3(\Theta)}{4F^2(\Theta)} \,, \hspace{15mm} F(\Theta)\equiv K(\Theta)K_3 (\Theta)\,,
\end{equation} 
can be straightforwardly integrated to obtain~\cite{Karananas:2016kyt}
\be{req_gho_free_2} 
K(\Theta)=-\frac{ M_P^2 K^{'2}_3(\Theta)}{4 \, K_3(\Theta)
( \kappa\,  K_3(\tilde \theta)+c)}\ ,
\ee
with $c$ an arbitrary constant. Using this expression and assuming $U$ and $K_\Lambda$ to be expressible in terms of $K_3$,  we can rewrite Eq.~\eqref{eq:sigmamodel} as
\begin{equation}\label{MSattractor}
S=\int d^4x \sqrt{-g}\, \Bigg\{\frac{M_P^2}{2}R -
\frac{1}{2}\Bigg[-\frac{M_P^2(\partial \bar \Theta)^{2}}{4\, \bar \Theta (\kappa
\bar \Theta+c)}+ \bar \Theta(\partial \Phi)^2 \Bigg] - U(\bar \Theta)-\Lambda_0\,
K_\Lambda(\bar \Theta)\,e^{-4 \Phi/M_P }\Bigg\}\,,
\end{equation}
where we have defined a variable $\bar \Theta\equiv K_3(\theta)$ to highlight the similarities  with the Higgs-Dilaton action \eqref{action_HD2}.  As in that case, the scale current attractor \eqref{eq:conserv} is generically active for sufficiently small values of $\Lambda_0$, reducing the system to a single field scenario in $\bar \Theta$. On top of that, the powerful pole structure in Eq.~\eqref{MSattractor} makes the inflationary dynamics almost insensitive to the details of the unspecified potential $U(\bar\Theta)$, generalizing the specific predictions \eqref{As} and \eqref{nsr2} to the full \textit{equivalence class} \eqref{MSattractor} \cite{Karananas:2016kyt, Casas:2018fum}. The pole structure of Eq.~\eqref{MSattractor} is indeed clearly reflected in these expressions.  For $4 c N_*\gg 1$, the spectral tilt approaches minus infinity while its running and the tensor-to-scalar ratio tend to zero, in good agreement with the single pole behavior at $c\neq 0$~\cite{Terada:2016nqg}. On the other hand, the emerging quadratic pole at $c\to 0$ makes the expressions converge to the well-known Higgs inflation results \cite{Rubio:2018ogq,Tenkanen:2020dge}
\begin{equation}\label{nsr2app}
A_s
 =\frac{\lambda_{s}N_*^2(1+6\alpha\kappa_c)^2}{12\pi^2|\kappa_c|}\,,  \hspace{10mm}
n_s \simeq 1-\frac{2}{N_*} \,,  \hspace{10mm} 
\alpha_s \simeq  - \frac{2}{N_*^2}\,, \hspace{10mm} r\simeq \frac{2}{\vert \kappa_c\vert N_*^2}\,,
\end{equation}
with  $\vert\kappa_c\vert\simeq 1/6$ in the metric case ($y=1$) and  $\kappa_c\simeq  \xi_{h}$ in the Palatini one ($y=0$). In this sense, the predictions of Higgs and Higgs-Dilaton inflation are not attached to specific metric or Palatini formulations, as sometimes stated in the literature, nor to the precise choice of the theory-defining functions in \eqref{eqn:hdm_lagrangianL}, but rather to a defining principle: \textit{the existence of an Einstein-frame target manifold with approximately constant curvature!}

\section*{Acknowledgements} 

I am glad to thank Mikhail Shaposhnikov, Fedor Bezrukov, Georgios Karananas, Martin Pauly and Santiago Casas for long-standing collaborations on the covered topics. 

\bibliographystyle{JHEP}
\bibliography{references}

\end{document}